\documentclass[12pt,a4paper]{article}

\usepackage{subfigure}
\usepackage{array}
\usepackage{graphicx}
\usepackage{amsmath}
\usepackage{amssymb}
\usepackage{mathrsfs}
\usepackage{bm}
\usepackage{color}
\usepackage{slashed}
\usepackage{threeparttable}
\usepackage[colorlinks, linkcolor=black, anchorcolor=black, citecolor=black]{hyperref}
\usepackage[amsmath,thmmarks,hyperref]{ntheorem}
\usepackage{soul}

\title{Two-Body Strong Decay of $Z(3930)$ as the $\chi_{c2} (2P)$ State}
\author{Tianhong Wang\footnote{thwang.hit@gmail.com}, Guo-Li Wang\footnote{gl\_wang@hit.edu.cn}, Hui-Feng Fu,  and Wan-Li Ju\\
{\it \small   Department of Physics, Harbin Institute of Technology,
Harbin, 150001, China} }
\date{\today}

\begin{document}
\maketitle

\begin{abstract}
The new particle $Z(3930)$ found by the Belle and BaBar Collaborations through the $\gamma\gamma\rightarrow D\bar D$ process is identified to be the $\chi_{c2}(2P)$ state. Since the mass of this particle is above the $D\bar D^{(\ast)}$ threshold, the OZI-allowed two-body strong decays are the main decay modes. In this paper, these strong decay modes are studied with two methods. One is the instantaneous Bethe-Salpeter method within Mandelstam formalism. The other is the combination of the $^3P_0$ model and the former formalism. The total decay widths are $26.3$ and $27.3$ MeV for the methods with or without the $^3P_0$ vertex, respectively. The ratio of $\Gamma_{D\bar D}$ over $\Gamma_{D\bar D^\ast}$ which changes along with the mass of the initial meson is also presented.
\end{abstract}

\section{Introduction}

It is well known that the quark potential models predict abundant charmonium spectra. In these spectra, for the $S$-wave, both the radial excitated spin-singlet and triplet states have been observed experimentally. For others such as the four $P$-wave states $^3P_J$ and $^1P_1$, all the ground states have been observed and studied carefully. The identification of $P$-wave radial excitations will greatly support the correctness of potential models and supply more playgrounds for the study of QCD non-perturbative properties.

Recently, the Belle Collaboration~\cite{belle1} found a new particle in the process $\gamma\gamma\rightarrow D\bar D$ with statistical significance of 5.3$\sigma$. The observed mass and decay width are $M=3929\pm5\pm2$ MeV and $\Gamma_{total}= 29\pm10\pm2$ MeV, respectively. Later through the same process while with a little higher statistical significance 5.8$\sigma$, the Babar Collaboration~\cite{babar1} also observed this particle with compatible mass and decay width: $M=3926.7\pm2.7\pm1.1$ MeV and $\Gamma_{total}= 21.3\pm6.8\pm3.6$ MeV. The experimental data favor the $2^{++}$ assignment, of which the (helicity) angular distribution of the decay products has a form $\sin^4\theta$. In Ref.~\cite{Liu}, Chen {\sl et al.} assumed that two $P$-wave higher charmonia participate as the intermediate state in the process, which is also compatible with the experimental data of two groups. However, as pointed by Ref.~\cite{Guo}, $X(3915)$ only plays a minor role in the data fitting of Ref.~\cite{Liu}. So a careful study of the decay properties of the $\chi_{c2}(2P)$ state is still needed, which is helpful for the further investigation of this state.

Since the mass of this particle is above the $D\bar D^{(\ast)}$ threshold, but under the $D^\ast\bar D^\ast$ threshold, the former channels are allowed by the Okubo-Zweig-Iizuka (OZI) rule and can be used to estimate the total decay width of this state. As is well known, these open-flavor strong decays closely relate to the non-perturbative properties, of which our knowledge is rather poor. A complete solution to this problem needs deep understanding of the QCD vacuum. Although we expect the lattice QCD calculations will provide firm theoretical predictions in the future, now we are forced to construct phenomenological models, e.g. the $^3P_0$ model~\cite{Micu, Le1, Le2}. Possible microscopic models include the flux-tube mode~\cite{KI}, Cornell model~\cite{Cornell1,Cornell2} with a vector confinement interaction, the model in Ref.~\cite{Swanson} with a scalar confinement interaction, Field Correlator Method~\cite{Simonov2} {\sl et al}.

Here we try to apply the Bethe-Salpeter (BS) method~\cite{BS1, BS2} to the two-body strong decay processes within Mandelstam formalism~\cite{Man}. It is well known that this method provides a relativistic description for the two-body bound state. Its instantaneous version is extensively used to study the properties of heavy mesons~\cite{wang, wang1, wang2}. One notices that this method has been mentioned in Ref.~\cite{book} and used to study light meson decays in Ref.~\cite{Ricken}, where in addition to the pure quark loop diagram, an instanton-induced vertex is considered for the processes with vanishing total angular momentum. Here for the charmonium case, we will assume the leading order diagram gives the main contribution. Non-perturbative effects are included in the wave functions of initial and final mesons. These wave functions are obtained by solving corresponding coupled instantaneous BS equations. As expected, the relativistic correction is covered, which is especially important for the orbital-excited particles.

In this method we make two approximations. First, we will not consider the interactions of final mesons, which can be realized by interchanging pions~\cite{Braaten}. This maybe important for the threshold-nearby decay processes. It is not clear now how to incorporate these effects into the BS method. Second, we will not consider the coupled-channel effects~\cite{Yu, Simonov1, Ferretti}, which may move the pole mass to the physical one. Here we adjust the parameter in the potential to get the correct mass value and obtain the wave function for the initial meson.

For comparison, we will also consider the $^3P_0$ model, which assumes that a $q\bar q$ pair is created from the vacuum with a quantum number $0^{++}$. This model is extensively applied to study the strong decays of light mesons~\cite{Swanson}, heavy-light mesons~\cite{Swanson1} and heavy quarkonia~\cite{Barnes}. It is a non-relativistic model, where the simple harmonic oscillator (SHO) wave functions are used both for the initial meson in its c.m. system and the final particles in their moving systems. As mentioned in Ref.~\cite{Simonov2}, this assumption is valid only for small relative velocity near the threshold. For $P$-wave charmonia, more covariant formalisms and realistic wave functions are needed if one wants to get a more reasonable result. In Ref.~\cite{Fu}, we have combined the BS method with the $^3P_0$ model to make it applicable for the OZI-allowed two-body decay processes. There we also proved that this amplitude will reduce to the usual form when the non-relativistic approximation is made. The same method will be considered in this paper.

The paper is organized as follows. In Section 2, we present the calculations by using the instantaneous BS method within Mandelstam formalism. The leading order amplitude is considered. In Section 3, within the same formalism, we add a phenomenological $^3P_0$ vertex to describe the decay mechanism. The results and discussions are given in Section 4.

\section{BS Method}

The BS equation is written as
\begin{equation}\label{BS}
\begin{aligned}
S^{-1}_1(p_1)\chi_{_P}(q)S^{-1}_2(-p_2)=i\int\frac{d^4k}{(2\pi)^4} V(P; q,k)\chi_{_P}(k),
\end{aligned}
\end{equation}
where $\chi_{_P}(q)$ is the BS wave function; $S_1(p_1)$ and $S_2(-p_2)$ are the quark and anti-quark propagators; $V(P; q,k)$ is the interaction kernel; $p_1$ and $p_2$ are the momenta of quark and anti-quark, respectively, which are related to the meson momentum $P$ and relative momentum $q$ by
\begin{equation}\label{momentum}
 p_{i} = \frac{m_{i}}{m_{1}+m_{2}}P+Jq.
\end{equation}
where $J=1$ for the quark ($i=1$) and $J=-1$ for the anti-quark ($i=2$).
With the definition $p_{i_P} \equiv \frac{P\cdot p_i}{M}$ and  $p^\mu_{i\perp} \equiv p_i^\mu - \frac{P\cdot p_i}{M^2}P^\mu$, we can write $S_i(Jp_i)$ as
\begin{equation}
\label{propagator}
\begin{aligned}
&-iJS_i(Jp_i)=\frac{\Lambda^+_i}{p_{i_P}-\omega_i+i\epsilon}+\frac{\Lambda_i^-}{p_{i_P}+\omega_i-i\epsilon},
\end{aligned}
\end{equation}
where we have used
\begin{equation}
\label{projector}
\begin{aligned}
&\Lambda_i^{\pm}(p^\mu_{i\perp})\equiv\frac{1}{2\omega_i}[\frac{\slashed P}{M}\omega_i\pm(\slashed p_{i\perp} + J m_i)],\\
&\omega_i \equiv \sqrt{m_i^2 - p_{i\perp}^2}.
\end{aligned}
\end{equation}

The 3-dimensional form of the BS wave function is defined as $\varphi(q^\mu_\perp)\equiv i\int\frac{dq_{_P}}{2\pi}\chi_{_P}(q)$. With instantaneous approximation, $V(P; q, k) \approx V(P; q_\perp, k_\perp)$, Eq.~(\ref{BS}) can be written as
\begin{equation}
\label{bswf}
\begin{aligned}
\chi_{_P}(q)=S_1(p_1)\eta_{_P}(q_\perp)S_2(-p_2),
\end{aligned}
\end{equation}
where
\begin{equation}
\label{eta}
\eta_{_P}(q_\perp)=\int\frac{d^3\vec k}{(2\pi)^3} V(q_\perp,k_\perp)\varphi_{_P}(k_\perp).
\end{equation}

Within Mandelstam formalism, the transition amplitude of two-body strong decays has the form~\cite{book},
\begin{equation}\label{noapprox}
\begin{aligned}
\langle P_1P_2|S|P\rangle_{BS}&=\int\frac{d^4q}{(2\pi)^4}\frac{d^4q_1}{(2\pi)^4}\frac{d^4q_2}{(2\pi)^{4}}\chi^{ab}_{_P}(q)\bar\chi^{cd}_{_{P_2}}(q_2)\bar\chi^{ef}_{_{P_1}}(q_1)\\
&\times K^{ab; cd; ef}(P, q; P_1, q_1; P_2, q_2),
\end{aligned}
\end{equation}
where $\bar\chi_{_P}(q)$ is defined as $\gamma^0\chi_{_P}^\dagger(q)\gamma^0$; $a\sim f$ are the Dirac indices and the color indices are suppressed. $K$ is the irreducible Green function which represents the `three-meson vertex'. In the leading order (see Fig.~1), every two mesons share a fermion propagator and $K$ has the following form,
\begin{equation}\label{kernel}
\begin{aligned}
K\sim S^{-1}_2(-p_2)\otimes S^{-1}_1(p_{21})\otimes S^{-1}_1(p_1).
\end{aligned}
\end{equation}
The meaning of $p_{ij}$ is explicit from the diagram and they are related to the meson momentum $P_i$ and relative momentum $q_i$ by the same relation as in Eq.~(\ref{momentum}).

In Fig.~1, one can see there is no interaction vertex, such as the $^3P_0$ type which we will consider in the following
section, to create the light $q\bar q$ pair. This can be understood as follows. The light $q\bar q$ pair can be created by a
soft gluon radiated by the charm or anti-charm quark, however, such interaction should be absorbed into the kernel of Eq.~(\ref{BS}) for the final mesons. This is also required by the irreducibility of $K$.

Inserting Eq.~(\ref{kernel}) into Eq.~(\ref{noapprox}), we get the transition amplitude to the leading order,
\begin{equation}\label{tr1}
\begin{aligned}
&\langle P_1P_2|S|P\rangle_{BS}=(2\pi)^4\delta^4(P-P_1-P_2)\mathcal M_{BS} \\
&=C_f \int\frac{d^4q}{(2\pi)^4}\frac{d^4q_1}{(2\pi)^4}\frac{d^4q_2}{(2\pi)^{4}}{\rm Tr}[\chi_{_P}(q)S^{-1}_2(-p_2)(2\pi)^4\delta^4(p_2-p_{22})\bar\chi_{_{P_2}}(q_2)\\
&~~~\times S^{-1}_1(p_{21})(2\pi)^4\delta^4(p_{21}+p_{12})\bar\chi_{_{P_1}}(q_1)S^{-1}_1(p_1)(2\pi)^4\delta^4(p_1-p_{11})]\\
&=(2\pi)^4\delta^4(P-P_1-P_2)C_f\int\frac{d^4q}{(2\pi)^4}{\rm Tr}[\chi_{_P}(q)S_2^{-1}(-p_2)\bar\chi_{_{P_2}}(q_2)S^{-1}_1(p_{21})\\
&~~~\times \bar\chi_{_{P_1}}(q_1)S^{-1}_1(p_1)],
\end{aligned}
\end{equation}
where $C_f = \frac{1}{\sqrt 3}$ is the color factor. The relative momenta of final mesons are related to that of the initial meson by $q_i = q + (-1)^{i+1}(\alpha_{i}P-\alpha_{ii}P_i)$, where
\begin{equation}
\alpha_i = \frac{m_i}{m_1+m_2}, \hspace{5mm} \alpha_{ii} = \frac{m_{ii}}{m_{i1}+m_{i2}},  \hspace{5mm} {\rm for} ~~i= 1, 2.
\end{equation}

\begin{figure}\label{ozi1}
\centering
\includegraphics[scale=0.9]{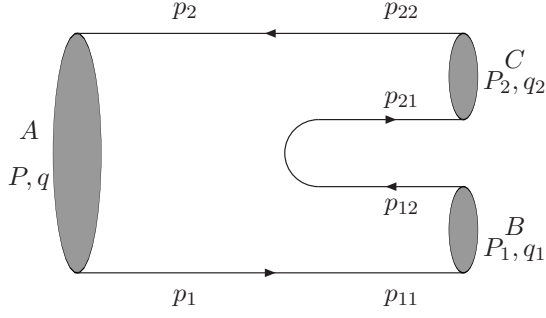}
\caption{Feynman diagram of OZI-allowed two-body decay process}
\end{figure}
\begin{figure}\label{ozi}
\centering
\includegraphics[scale=0.9]{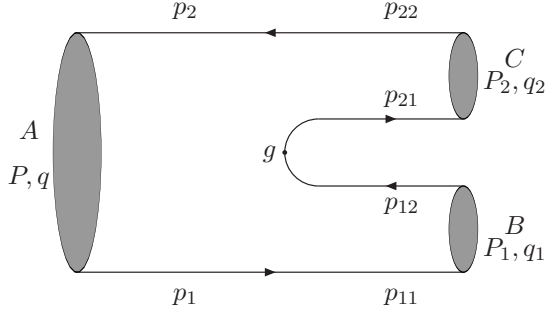}
\caption{Feynman diagram of OZI-allowed two-body decay process with a $^3P_0$ vertex.}
\end{figure}

From the Eq.~(\ref{tr1}), we get the Feynman amplitude,
\begin{equation}\label{LO}
\begin{aligned}
\mathcal M_{BS} &= C_f\int\frac{d^4q}{(2\pi)^4}{\rm Tr}[\chi_{_P}(q)S_2^{-1}(-p_2)\bar\chi_{_{P_2}}(q_2)S^{-1}_1(p_{21})\bar\chi_{_{P_1}}(q_1)S^{-1}_1(p_1)]\\
&=C_f\int\frac{d^4q}{(2\pi)^4}{\rm Tr}[S_1(p_{1})\eta_{_P}S_2(-p_{22})\bar\eta_{_{P_2}}S_2(-p_{12})\bar\eta_{_{P_1}}]\\
&\simeq -iC_f\int\frac{d^4q}{(2\pi)^4}{\rm Tr}[\frac{\Lambda_{1}^+(q_\perp)}{p_{1_{P}}-\omega_{1}+i\epsilon}\eta_{_P}\frac{\Lambda_{2}^+(q_\perp)}{p_{2_{P}}-\omega_{2}+i\epsilon}\bar\eta_{_{P_2}}(-\frac{\Lambda_{21}^+(q_{2\perp})}{p_{21_{P}}-\omega_{21}+i\epsilon}\\
&~~~~+\frac{\Lambda_{12}^+(q_{1\perp})}{p_{12_{P}}-\omega_{12}+i\epsilon})\bar\eta_{_{P_1}}]\\
&=C_f\int\frac{d^3\vec q}{(2\pi)^3}{\rm Tr}[\varphi_{_P}^{++}(q_\perp)\frac{\slashed P}{M}\overline{\varphi_{_{P_2}}^{++}}(q_{2\perp})\bar\eta_{_{P_1}}(q_{1\perp})-\varphi_{_P}^{++}(q_\perp)\bar\eta_{_{P_2}}(q_{2\perp})\\
&~~~~\times\overline{\varphi_{_{P_1}}^{++}}(q_{1\perp})\frac{\slashed P}{M}]\\
&=C_f\int\frac{d^3\vec q d^3\vec k}{(2\pi)^6}{\rm Tr}[\varphi_{_P}^{++}(q_\perp)\frac{\slashed P}{M}\overline{\varphi_{_{P_2}}^{++}}(q_{2\perp})\overline{V(q_{1\perp}- k_\perp)\varphi_{_{P_1}}(k_\perp)}\\
&~~~~- \varphi_{_P}^{++}(q_\perp)\overline{V(q_{2\perp}- k_\perp)\varphi_{_{P_2}}(k_\perp)}\overline{\varphi_{_{P_1}}^{++}}(q_{1\perp})\frac{\slashed P}{M}].
\end{aligned}
\end{equation}
In the second equation, we have used Eq.~(\ref{bswf}), and in the third equation, we have inserted Eq.~(\ref{propagator}). For simplicity, we only keep the positive energy projectors which give the main contributions. In the parentheses we have used $\Lambda^-_{12}(q_{1\perp}) = \Lambda^+_{21}(q_{2\perp})$. Considering the BS equation~\cite{wang}
\begin{equation}
\label{salpeter}
\begin{aligned}
&(M-\omega_1-\omega_2)\varphi^{++}_{_P}(q_\perp) = \Lambda_1^+\eta_{_P}(q_\perp)\Lambda_2^+,
\end{aligned}
\end{equation}
($\varphi^{++}$ is defined as $\Lambda^+_1\frac{\slashed P}{M}\varphi\frac{\slashed P}{M}\Lambda^+_2$) and by doing the residual integral, we get the fourth equation. In the fifth equation we have used Eq.~(\ref{eta}).

The wave functions of $2^+$, $1^-$ and $0^-$ states have the following forms~\cite{wang2, wang, wang1},
\begin{equation}\label{wf}
\begin{aligned}
&\varphi_{_{2^+}}=\epsilon_{\mu\nu}q^\mu[q^\nu(f_1+f_2\frac{\slashed P}{M}+f_3\frac{\slashed q}{M}+f_4\frac{\slashed P\slashed q}{M^2})+M\gamma^\nu(f_5+f_6\frac{\slashed P}{M}\\
&~~~~~~~~+f_7\frac{\slashed q}{M})+\frac{i}{M}f_8\epsilon^{\mu\alpha\beta\gamma} P_\alpha q_{\perp\beta}\gamma_\gamma\gamma_5],\\
&\varphi_{_{0^-}}=M_1(g_1\frac{\slashed P_1}{M_1}+g_2+g_3\frac{\slashed q_{1\perp}}{M_1}+g_4\frac{\slashed P_1\slashed q_{1\perp}}{M_1^2})\gamma^5,\\
&\varphi_{_{1^-}}=q_{2\perp}\cdot\epsilon[h_1+h_2\frac{\slashed P_2}{M_2}+h_3\frac{\slashed q_{2\perp}}{M_2}+h_4\frac{\slashed P_2\slashed q_{2\perp}}{M_2^2}]+M_2{\slashed\epsilon}(h_5+h_6\frac{\slashed P_2}{M_2})\\
&~~~~~~~~+(\slashed q_{2\perp}\slashed\epsilon - q_{2\perp}\cdot\epsilon)h_7+\frac{1}{M_2}(\slashed P_2\slashed\epsilon\slashed q_{2\perp} - \slashed P_2q_{2\perp}\cdot\epsilon)h_8,\\
\end{aligned}
\end{equation}
where $f_i$, $g_i$ and $h_i$ are functions of $\vec q$, $\vec q_1$ and $\vec q_2$, respectively.

To perform the integral in Eq.~(\ref{LO}), we have chosen $\vec P_1$ as the reference direction,
\begin{equation}
\begin{aligned}
\vec P_1\cdot \vec q =& |\bar P_1||\vec q|\cos\theta_1,~~~~  \vec P_1\cdot \vec k = |\bar P_1||\vec k|\cos\theta_2,\\
\vec q\cdot\vec k =& |\vec q||\vec k|\cos\theta,\\
\cos\theta=&\sin\theta_1\sin\theta_2\cos(\phi_1-\phi_2)+\cos\theta_1\cos\theta_2.
\end{aligned}
\end{equation}

By defining $P^\mu_{1\perp} = P^\mu_1 - \frac{P\cdot P_1}{M^2}P^\mu$ and $g^{\mu\nu}_{\perp} = g^{\mu\nu} - \frac{P^\mu P^\nu}{M^2}$, we can express the integrals which have free Lorentz indexes as follows,
\begin{equation}\label{integral}
\begin{aligned}
&\int d^3\vec q d^3\vec k q^\mu_\perp F(q_\perp, k_{\perp}) = f_{11} P^\mu_{1\perp},\hspace{1cm} \int d^3\vec q d^3\vec k k^\mu_{\perp} F(q_\perp, k_{\perp}) = g_{11} P^\mu_{1\perp},\\
&\int d^3\vec q d^3\vec k q^\mu_\perp q^\nu_\perp F(q_\perp, k_{\perp}) = f_{21} P^\mu_{1\perp} P^\nu_{1\perp} +f_{22} g^{\mu\nu}_\perp,\\
&\int d^3\vec q d^3\vec k k^\mu_{\perp} q^\nu_{\perp} F(q_\perp, k_{\perp}) = g_{21} P^\mu_{1\perp} P^\nu_{1\perp} +g_{22} g^{\mu\nu}_\perp,\\
&\int d^3\vec q d^3\vec k q^\mu_\perp q^\nu_{\perp} q^\alpha_{\perp} F(q_\perp, k_{\perp}) = f_{31} P^\mu_{1\perp} P^\nu_{1\perp} P^\alpha_{1\perp} +f_{32} (P^\mu_{1\perp} g^{\nu\alpha}_\perp+P^\nu_{1\perp} g^{\mu\alpha}_\perp+P^\alpha_{1\perp} g^{\mu\nu}_\perp),\\
&\int d^3\vec q d^3\vec k k^\mu_{\perp} q^\nu_{\perp} q^\alpha_{\perp} F(q_\perp, k_{\perp}) = g_{31} P^\mu_{1\perp} P^\nu_{1\perp} P^\alpha_{1\perp} +g_{32} P^\mu_{1\perp} g^{\nu\alpha}_\perp+g_{33}(P^\nu_{1\perp} g^{\mu\alpha}_\perp+P^\alpha_{1\perp} g^{\mu\nu}_\perp),
\end{aligned}
\end{equation}
where $f_{ij}$ and $g_{ij}$ are integrals with no free Lorentz indexes, which can be done numerically.

Inserting Eq.~(\ref{wf}) into Eq.~(\ref{LO}) and finishing the trace, we can get the transition amplitude with the following forms,
\begin{equation}\label{amplitude}
\begin{aligned}
&\mathcal M(\chi_{c2}(2P)\rightarrow D\bar D) = \epsilon_{\mu\nu}P_1^\mu P_1^\nu t_1,\\
&\mathcal M(\chi_{c2}(2P)\rightarrow D\bar D^\ast) = \epsilon^{\mu\alpha\beta\delta}\epsilon_{\mu\nu}\epsilon_{\alpha}P_1^\nu P_{\beta} P_{1\delta} t_2,
\end{aligned}
\end{equation}
where $t_1$ and $t_2$ are form factors, which are the functions of $f_{ij}$ and $g_{ij}$ in Eq.~(\ref{integral}).
These amplitudes actually can be constructed by using the momenta and polarization tensor (vector) of initial and final mesons. With consideration of parity properties, we can see there is a totally antisymmetric tensor in the second amplitude while the first one does not include it.

The decay width is
\begin{equation}
\begin{aligned}
\Gamma=\frac{|\vec P_1|}{8\pi M^2}\frac{1}{5}\sum_\lambda|\mathcal M|^2,
\end{aligned}
\end{equation}
where $|\vec P_1| = \sqrt{[M-(M_1-M_2)^2][M-(M_1 + M_2)^2]}/2M$ is the final meson momentum.

\section{Extended $^3P_0$ Model}
\label{E3P0}

The usual $^3P_0$ model is a non-relativistic model with a transition operator $g\int d\vec{x}\bar{\psi}\psi|_{\mathrm{nonrel}}$~\cite{Swanson2}. To combine it with BS wave functions, we extend it to the relativistic form $-ig\int d^4 x\bar\psi\psi$
(a similar form of interaction is also used in Refs.~\cite{Simonov2, Simonov1}), where $g$ is parameterized as $2m_q\gamma$. $m_q$ is the constitute quark ($u$, $d$, $s$) mass. The interaction strength $\gamma$ is dimensionless, which is roughly flavor independent in the light meson decay processes~\cite{Barnes}. Just as Ref.~\cite{Barnes} did, we will assume this also applies in the heavy meson case.

Here for simplicity we will not get the value of $\gamma$ by fitting decay widths of other channels, instead, we take $\gamma=0.35$, which is the best-fit value for the usual $^3P_0$ model~\cite{Swanson2}. In this case, we can only get a roughly estimation for our extended $^3P_0$ model. But considering that this model will reduce to the usual one if non-relativistic approximation is made, we can argue that by using the same interaction strength we get qualitatively how large change can be brought by using a covariant formalism and BS wave functions. A more reliable quantity is the ratio of partial decay widths of different channels.

Within Mandelstam formalism we can write the transition amplitude of the OZI-allowed open flavor decay process (see Fig.~\ref{ozi}) as
\begin{equation}
\begin{aligned}
&\langle P_1P_2|S|P\rangle_{^3P_0}=(2\pi)^4\delta^4(P-P_1-P_2)\mathcal M_{^3P_0} \\
&= -ig\int\frac{d^4q}{(2\pi)^4}\frac{d^4q_1}{(2\pi)^4}\frac{d^4q_2}{(2\pi)^{4}}{\rm Tr}[\chi_{_P}(q)S^{-1}_2(p_2)(2\pi)^4\delta^4(p_2-p_{22})\bar\chi_{_{P_2}}(q_2)\bar\chi_{q_1}\\
&~~~~\times S^{-1}_1(p_1)(2\pi)^4\delta^4(p_1-p_{11})]\\
&= -ig(2\pi)^4\delta^4(P-P_1-P_2)\int\frac{d^4q}{(2\pi)^4}{\rm Tr}[\chi_{_P}(q)S_2^{-1}(-p_2)\bar\chi_{_{P_2}}(q_2)\bar\chi_{_{P_1}}(q_1)S_1^{-1}(p_1)]
\end{aligned}
\end{equation}
From the above equation, we get the Feynman amplitude~\cite{Fu}
\begin{equation}
\begin{aligned}
\mathcal M_{^3P_0}&=-ig\int\frac{d^4q}{(2\pi)^4}{\rm Tr}[\chi_{_P}(q)S_2^{-1}(-p_2)\bar\chi_{_{P_2}}(q_2)\bar\chi_{_{P_1}}(q_1)S_1^{-1}(p_1)]\\
&=g\int\frac{d^3\vec q}{(2\pi)^3} {\rm Tr} [\frac{\slashed P}{M}\varphi^{++}_{_P}(q_\perp)\frac{\slashed P}{M}\overline\varphi^{++}_{_{P_2}}(q_{2\perp})\overline\varphi^{++}_{_{P_1}}(q_{1\perp})](1-\frac{M-\omega_1-\omega_2}{2\omega_{12}})
\end{aligned}
\end{equation}
where $q_i = q + (-1)^{i+1}(\alpha_{i}P-\alpha_{ii}P_i)$. To get the second equation, we have made the residual integral and used Eq.~(\ref{salpeter}).
The second term in the last parentheses can be neglected for $M\approx\omega_1+\omega_2$ (for large $|\vec q|$, this approximation is not valid, while at the same time the wave function is strongly suppressed). Inserting Eq.~(\ref{wf}) into above equation we get the same forms of the decay amplitudes as Eq.~(\ref{amplitude}).

\section{Results and Discussions}
\label{Discuss}
\begin{table}
\caption{The value of $-V_0$ (GeV) for different states.}
\label{v0}
\setlength{\tabcolsep}{0.05cm}
\centering
\begin{tabular*}{\textwidth}{@{}@{\extracolsep{\fill}}ccccccc}
\hline\hline
States&$\chi_{c2}(2P)$&$D^+$ &$D^0$&$D_s^+$&$D^{\ast+}$&$D^{\ast 0}$\\ \hline
{\phantom{\Large{l}}}\raisebox{+.2cm}{\phantom{\Large{j}}}
$-V_0$&$0.177\sim0.098$&0.375&0.375 &0.432&0.11&0.11\\
\hline\hline
\end{tabular*}
\end{table}

\begin{table}
\caption{OZI-allowed two-body strong decay widths (MeV) of $\chi_{c2}(2P)$ with two methods, where we have used $M=3930$ MeV. The uncertainties are given by varying all the input parameters simultaneously within $\pm 5$\%.}
\label{patial}
\setlength{\tabcolsep}{0.05cm}
\centering
\begin{tabular*}{\textwidth}{@{}@{\extracolsep{\fill}}ccccccc}
\hline\hline
Mode&$D^+D^-$ &$D^0\bar D^0$&$D\bar D$&$D^+D^{\ast-}$&$D^0\bar D^{\ast 0}$&$ D\bar D^\ast$\\ \hline
{\phantom{\Large{l}}}\raisebox{+.2cm}{\phantom{\Large{j}}}
$\chi_{c2}(2P)$ (BS)&$8.57^{+4.24}_{-1.76}$&$9.47^{+4.48}_{-1.93}$&$18.0^{+8.72}_{-3.69}$ &$1.84^{+0.62}_{-0.52}$&$2.33^{+0.87}_{-0.61}$&$8.34^{+2.98}_{-2.26}$\\
{\phantom{\Large{l}}}\raisebox{+.2cm}{\phantom{\Large{j}}}
$\chi_{c2}(2P)$ (BS-$^3P_0$)&$10.1^{+2.6}_{-2.3}$&$10.3^{+2.5}_{-2.2}$&$20.4^{+5.1}_{-4.5}$&$1.48^{+0.64}_{-0.39}$&$1.97^{+0.75}_{-0.48}$&$6.90^{+2.78}_{-1.74}$ \\
\hline\hline
\end{tabular*}
\end{table}

\begin{figure}\label{width}
\centering
\subfigure[]{\includegraphics[scale=0.77]{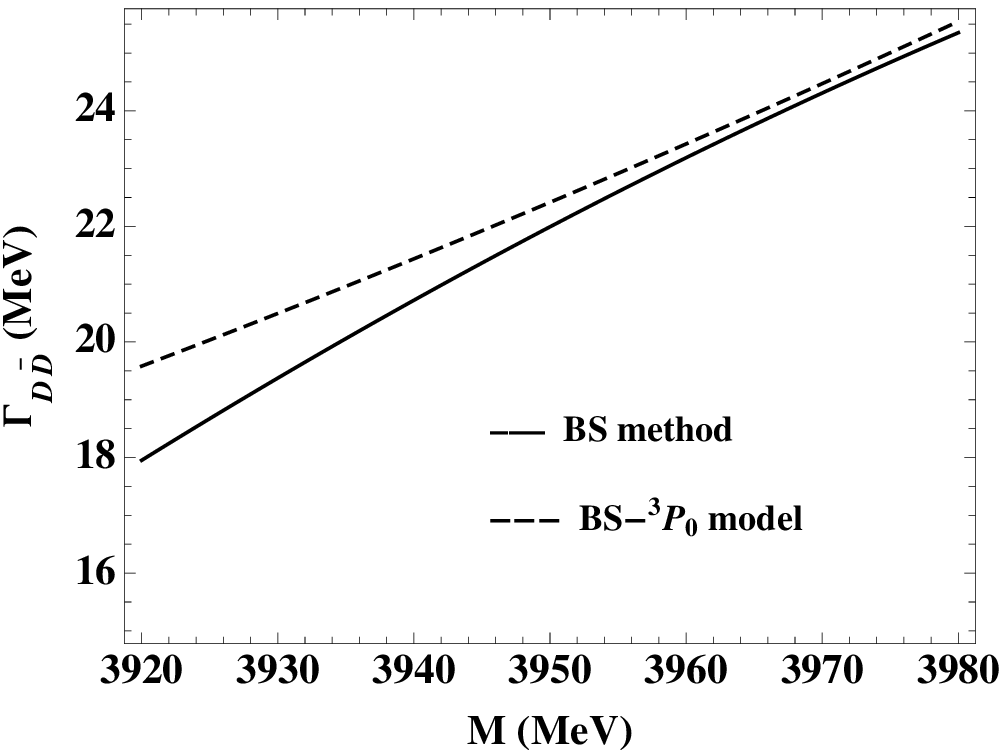}}
\subfigure[]{\includegraphics[scale=0.77]{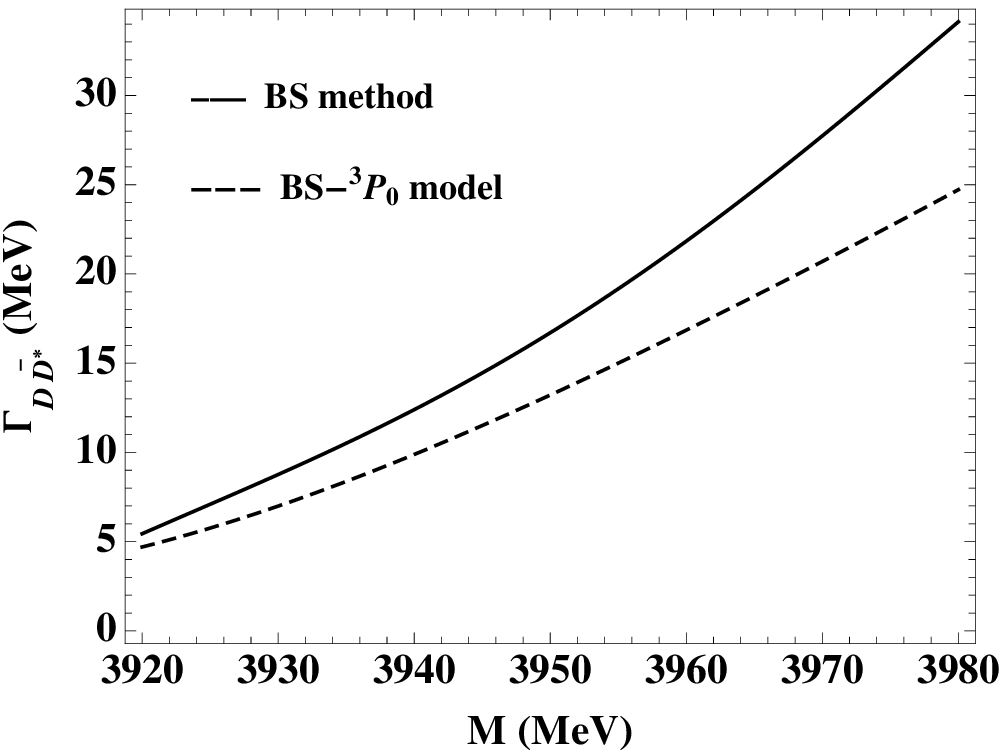}}
\subfigure[]{\includegraphics[scale=0.78]{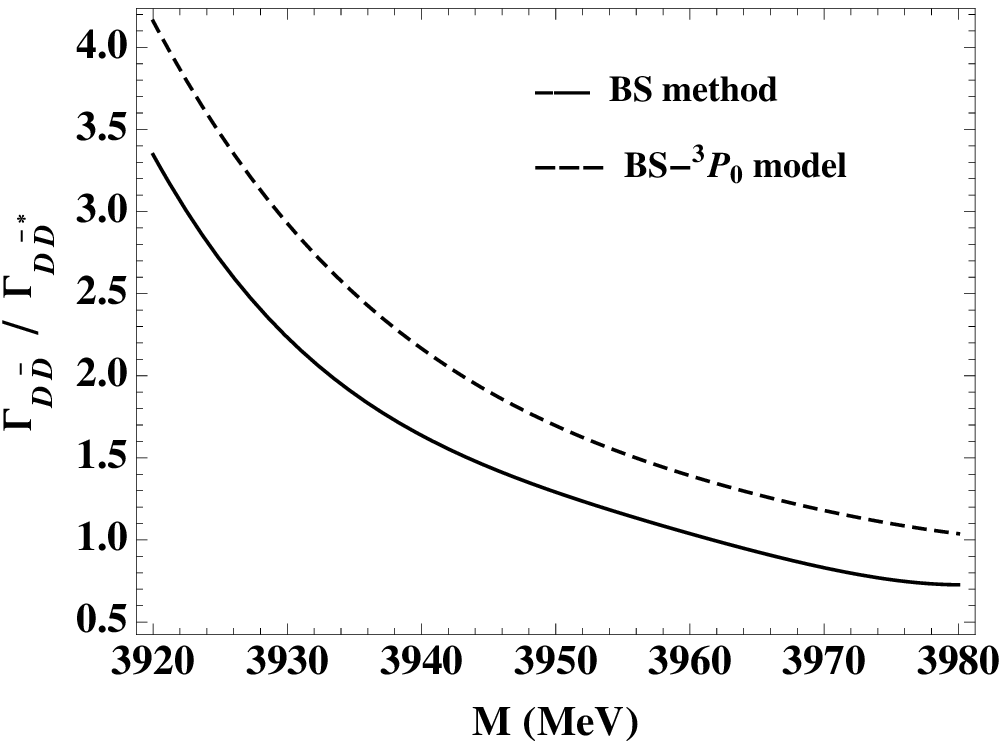}}
\caption{(a) $\Gamma_{D\bar D}$, (b) $\Gamma_{D\bar D^{\ast}}$ and (c) $\Gamma_{D\bar D}/\Gamma_{D\bar D^{\ast}}$ change along with the mass of the initial meson $\chi_{c2}(2P)$. Here we only give the results without uncertainties. If the uncertainties are considered, the two lines will overlap each other.}
\end{figure}

\begin{table}
\caption{Open-flavor strong decay widths (MeV) of $\chi_{c2}(2P)$ state. The second and the third columns are our results with BS method and BS-$^3P_0$ model, respectively, where the values outside the parentheses are gotten with $M=3930$ MeV, while the results in the parentheses are calculated with $M=3972$ MeV. The uncertainties are given by varying all the input parameters simultaneously within $\pm 5$\%. The fourth column is the usual $^3P_0$ model and in the fifth column, ${\rm C}^3$ represents the Cornell coupled-channel model.}
\label{results}
\setlength{\tabcolsep}{0.1cm}
\centering
\begin{threeparttable}
\begin{tabular*}{\textwidth}{@{}@{\extracolsep{\fill}}ccccccc}
\hline\hline
Mode&BS&BS-$^3P_0$ &$^3P_0$~\cite{Swanson2}\tnote{$\ast$}&${\rm C}^3$~\cite{Eichten}&Ref.~\cite{Liu}&Exp~\cite{PDG}\\ \hline
{\phantom{\Large{l}}}\raisebox{+.2cm}{\phantom{\Large{j}}}
$D\bar D$&$18.0^{+8.7}_{-3.7}$ ($24.8^{+8.6}_{-6.2}$)& $20.4^{+5.1}_{-4.9}$ ($24.6^{+4.9}_{-4.5}$)&26 (32)&21.5&&\\
{\phantom{\Large{l}}}\raisebox{+.2cm}{\phantom{\Large{j}}}
$D_s\bar D_s$&--- ($0.347^{+0.291}_{-0.216}$)&--- ($0.620^{+0.218}_{-0.162}$) &--- (0.5)&---&& \\
{\phantom{\Large{l}}}\raisebox{+.2cm}{\phantom{\Large{j}}}
$D\bar D^{\ast}$&$8.34^{+2.98}_{-2.26}$ ($34.7^{+8.0}_{-7.9}$)&$6.90^{+2.78}_{-1.74}$ ($21.1^{+7.5}_{-5.6}$) &9 (28)&7.1&&\\
{\phantom{\Large{l}}}\raisebox{+.2cm}{\phantom{\Large{j}}}
$total$&$26.3^{+11.7}_{-6.0}$ ($59.8^{+16.9}_{-14.3}$)&$27.3^{+7.9}_{-6.6}$ ($46.3^{+12.6}_{-10.3}$) &35 (60.5)&28.6&11.98&$24\pm6$\\
\hline\hline
\end{tabular*}
	\begin{tablenotes}
		\footnotesize
			\item[$\ast$] In Ref.~\cite{Barnes}, the authors used $\gamma$ = 0.4 for the $^3P_0$ model with $M$ = 3972 MeV. There they got $\Gamma_{total}$ = 80 MeV.
	\end{tablenotes}
\end{threeparttable}
\end{table}

The potential in Eq.~(\ref{LO}) is the Cornell potential. It is the same one as we used to solve the instantaneous BS equations. The parameters in the potential are fixed by fitting the mass spectra. So we do not introduce new parameters when calculate the decay width. In the momentum space, it has the form
\begin{equation}
\begin{aligned}
\label{Cornell}
V(\vec{q})=(2\pi)^3V_s(\vec{q})
+\gamma_0\otimes\gamma^0 (2\pi)^3 V_v(\vec{q}),\\
V_{s}(\vec{q})
=-(\frac{\lambda}{\alpha}+V_0)\delta^{3}(\vec{q})
+\frac{\lambda}{\pi^{2}}\frac{1}{(\vec{q}^{2}+\alpha^{2})^{2}},\\
V_v(\vec{q})=-\frac{2}{3\pi^{2}}
\frac{\alpha_{s}(\vec{q})}{\vec{q}^{2}+\alpha^{2}},\\
\alpha_s(\vec{q})=\frac{12\pi}{27}
\frac{1}{{\rm{ln}}(a+\frac{\vec{q}^2}{\Lambda_{QCD}})},
\end{aligned}
\end{equation}
where the following parameter values are used: $a=e=2.7183$, $\alpha$ = 0.06
GeV, $\lambda$ = 0.21 ${\rm GeV}^2$, $m_c$ = 1.62 GeV, $m_s$ = 0.5 GeV, $m_u$ = 0.305 GeV, $m_d$ = 0.311 GeV, $\Lambda_{QCD}$ = 0.27 GeV. The value of $-V_0$ is listed in Table~\ref{v0}. For $\chi_{c2}(2P)$, we vary it in the range: 0.177$\sim$0.098 GeV. If we let the mass of $\chi_{c2}(1P)$ coincident with the experimental result, we get $-V_0=0.11$ GeV and $M(\chi_{c2}(2P))=3972$ MeV, which is about 40 MeV larger than the experimental data. To get $M(\chi_{c2}(2P))=3930$ MeV, we scale $-V_0$ to 0.165 GeV.

The patial decay widths for $\chi_{c2}(2P)$ ($M=3930$ MeV) with two methods are given in Table~\ref{patial}, where $D\bar D$ represents $D^+D^-+D^0\bar D^0$ and $D\bar D^\ast$ represents $D^+D^{\ast-}+D^0\bar D^{\ast0} + c.c.$. One notices that the $D\bar D^\ast$ channel contributes about half (a third) of that of the $D\bar D$ channel for the BS (BS-$^3P_0$) model. A precise measurement by future experiments for the ratio of two partial widths is expected. The ratio $Br(\chi_{c2}(2P)\rightarrow D^+D^-)/Br(\chi_{c2}(2P)\rightarrow D^0\bar D^0)$ is 0.90 (for BS), 0.98 (for BS-$^3P_0$), which is consistent with the experimental value~\cite{belle1}: 0.74 $\pm$ 0.43 (stat) $\pm$ 0.16 (syst). For the $D\bar D^\ast$ case, $Br(\chi_{c2}(2P)\rightarrow D^+D^{\ast-})/Br(\chi_{c2}(2P)\rightarrow D^0\bar D^{\ast0})$ = 0.79 is smaller than that of $D\bar D$. This is because the former channel has smaller phase space, a small mass difference $M(D^+D^{\ast-})-M(D^0\bar D^{\ast0})$ will cause a large difference of the partial decay widths.

In Table~\ref{results}, we present results with different models and experimental data. Our two methods get quite close total decay widths, which are also consistent with the experimental value. For partial decay widths, BS-$^3P_0$ model gives a larger $\Gamma_{D\bar D}$, while for $\Gamma_{D\bar D^\ast}$, BS model gets the larger results. By using the usual $^3P_0$ model, Ref.~\cite{Swanson2} get a large result compared with our BS-$^3P_0$ model. As mentioned in Section~\ref{E3P0}, for we have adopted the same interaction strength, the discrepancy may come from using different formalisms and wave functions. In Ref.~\cite{Eichten}, the Cornell coupled-channel (${\rm C}^3$) model is used, which assumes a current-current interaction with a vector confinement. By using gaussian type wave functions and solving the coupled-channel equation, the authors get the complex eigenvalue whose real part is the physical mass of the meson and the imaginary part related to the decay width. One notices this model gives results very close to ours.

In Fig.~3, we give the partial decay widths for $D\bar D$ and $D\bar D^\ast$ channels and the ratio of the two widths which changes along with the initial meson mass. One can see for $\Gamma_{D\bar D}$, the BS-$^3P_0$ model gives larger results and the difference of two models becomes less if the initial meson mass increases. For $\Gamma_{D\bar D^\ast}$, the BS model gives larger values and the difference of two models becomes larger along with the increasing of $\chi_{c2}(2P)$ mass. The BS-$^3P_0$ model gives a larger ratio of two partial widths, which decreases about 4 times when the mass of the $\chi_{c2}(2P)$ state increases from 3920 MeV to 3980 MeV. One notices that the partial widths with the $^3P_0$ model are $\gamma$-dependent, while the ratio is $\gamma$-independent and more reliable.

By using the widths in Table~\ref{results}, we can calculate the branching ratio $\Gamma_{D\bar D}/\Gamma_{total}$ = 0.684 (BS) and 0.747 (BS-$^3P_0$). In Ref.~\cite{wang2}, the same method gives $\Gamma_{\gamma\gamma}=0.534$ keV for $\chi_{c2}(2P)$. Combining the two results, we get $\Gamma_{\gamma\gamma}Br[\chi_{c2}(3930)\rightarrow D\bar D] = 0.365$ keV (BS) and 0.399 keV (BS-$^3P_0$). The experimental data are~\cite{belle1, babar1}:
$0.18\pm0.05\pm0.03$ keV (Belle) and
$0.24\pm0.05\pm0.04$ keV (BaBar). The theoretical results are consistent with the experiment values considering the large uncertainty of the QCD corrections to the diphoton decay processes \cite{kwong}.

In conclusion, we have used two methods to calculate the OZI-allowed two-body strong decay processes of the $\chi_{c2}(2P)$ state: the BS method and the extended $^3P_0$ model. In the former we did not introduce any new parameters, while in the later a flavor-independent interaction strength is used. The total decay width estimated is 26.3 (27.3) MeV for the former (later) model with $M=3930$ MeV, which is consistent with the experimental data. However, more experiments, especially the precise measurements of the partial decay widths of $D\bar D$ and $D\bar D^\ast$ channels are still needed for the final confirmation of this paritcle.

\section{Acknowledgments}

This work was supported in part by the National Natural Science
Foundation of China (NSFC) under Grant No.~11175051.

\end{document}